\documentclass[11pt]{article}
\usepackage{mydef2col}
\usepackage{kantlipsum,widetext}
\usepackage[T1]{fontenc}
\usepackage[autostyle=true]{csquotes}

\allowdisplaybreaks[3]
\usepackage{tikz}
\usetikzlibrary{calc,decorations.markings}
\usetikzlibrary{arrows,patterns,positioning}

\usepackage[round]{natbib}
\def\baru{{\bar{u}}}

\newcommand{\iu}{\mathrm{i}\mkern1mu}

\title{Semi-closed form prices of barrier options  in the Hull-White model}
\def\thetitle1{Semi-closed form solutions for barrier options ...}
\author{
\authorstyle{
Andrey Itkin{}
\textsuperscript{1}
and Dmitry Muravey
\textsuperscript{2}
}
\newline\newline
\textsuperscript{1}
\institution{Tandon School of Engineering, New York University, 1 Metro Tech Center, 10th floor, Brooklyn NY 11201, USA} \\
\textsuperscript{2}
\institution{Moscow State University, Moscow, Russia}
}

\date{\today}
\begin{document}

\maketitle

\lettrineabstract{In this paper we derive semi-closed form prices of barrier (perhaps, time-dependent) options for the Hull-White model, ie., where the  underlying follows a time-dependent OU process with a mean-reverting drift. Our approach is similar to that in (Carr and Itkin, 2020) where the method of generalized integral transform is applied to pricing barrier options in the time-dependent OU model, but extends it to an infinite domain (which is an unsolved problem yet). Alternatively, we use the method of heat potentials for solving the same problems.  By semi-closed solution we mean that first, we need to solve numerically a linear Volterra equation of the first kind, and then  the option price is represented as a one-dimensional integral. Our analysis shows that computationally our method is more efficient than the backward and even forward finite difference methods (if one uses them to solve those problems), while providing better accuracy and stability.}

%%%%%%%%%%%%%%%%%%%%%%%%%%%%%%%%%%%%%%%%%%%%%%%%%%%%%%%%%%%%
\vspace{0.5in}

\section*{Introduction}

The Hull-White model since it was invented in \citep{Hull:1990a} became to be very popular among practitioners for modeling interest rates and credit. That is because it is relatively simple and allows for negativity prices (while for a long time this behaviour was treated as a deficiency of the model, nowadays this became its advantage). The model could be calibrated to the given term-structure of interest rates and to the prices or implied volatilities of caps, floors or European swaptions since the mean-reversion level and volatility are functions of time. Under the Hull-White model the prices of Zero-coupon bonds and European Vanilla options are known in closed form, \citep{andersen2010interest}. However, for exotic options, e.g., barrier options, these prices are not known yet in closed form. Therefore, various numerical methods are used to obtain such prices, see survey in \citep{KuanWebber2001}.

In this paper we present an analytical solution of this problem, and demonstrate that it significantly accelerates computation of the prices and, accordingly, calibration of the model. Our contribution is twofold. First, we solve the problem of pricing Barrier options in semi-closed form and provide the resulting expressions not known yet in the literature. Second, we solve this problem by two methods. The first one is the method of heat potentials (HP), which came to mathematical finance due to A.~Lipton who borrowed it from mathematical physics (see references in the next section). The other method is a method of generalized integral transform (GIT), also known in physics, and introduced to mathematical  finance in \citep{CarrItkin2020}. However, this method solves the problem where the underlying is defined at the domain $S \in [0,y(t)]$ with $S$ being the stock price, and $y(t)$ being the time-dependent barrier. Contrary, in this paper we consider a complimentary domain $r \in [y(t), \infty)$ where the solution is not known yet. Therefore, our paper fills this gap, and the constructed solution can be applied to a wide class of problems, both in physics and finance.

\section{The model} \label{SecBO}

We consider a one-factor short interest rate model, first introduced in \citep{Hull:1990a}, and named after the authors as the Hull-White model. The model assumes dynamics of the short interest rate $r_t$ to follow the Ornstein-Uhlenbeck (OU) process with time-dependent coefficients
\begin{equation} \label{OU1}
d r_t = \kappa(t)[\theta(t) - r_t] dt + \sigma(t)dW_t, \qquad r_{t=0} = r.
\end{equation}
Here $t \ge 0$ is the time,  $\kappa(t) > 0$ is the constant speed of mean-reversion, $\theta(t)$ is the mean-reversion level, $\sigma(t)$ is the volatility of the process, $W_t$ is  the standard Brownian motion under the risk-neutral measure. To address calibration to real market rate curves the model could be updated by using a deterministic shift $s(t)$, so $r(t) = s(t) + \bar{r}(t)$ where $\bar(t)$ solves \eqref{OU1}. This can be easily done using our framework, see \citep{ItkinLiptonMuravey2020}, so here we don't concentrate on that.

This model is also popular for modeling prices of the plain vanilla and exotic options. In particular, in this paper we consider a Down-and-Out barrier option with the time-dependent lower barrier $L(t)$ where the underlying is a zero-coupon bond with maturity $S$ and price $F(r,t,S)$.

\subsection{The underlying - Zero Coupon Bond} \label{secZCB}

We assume that the short interest rate evolves in time as in \eqref{OU1}. It is known that $F(r,t,S)$ under a risk-neutral measure solves a linear parabolic partial differential equation (PDE), \citep{privault2012elementary}
\begin{equation} \label{PDE}
\fp{F}{t} + \dfrac{1}{2}\sigma^2(t) \sop{F}{r} + \kappa(t) [\theta(t) - r] \fp{F}{r} = r F.
\end{equation}
This equation should be solved subject to the terminal condition
\begin{equation} \label{termZCB}
 F(r,S,S)  = 1,
\end{equation}
\noindent and the boundary conditions
\begin{equation} \label{bc1}
 F(0,t,S)  = g(t), \qquad  F(r,t,S)\Big|_{r \to \infty} = 0,
\end{equation}
\noindent where $g(t)$ is some function of the time $t$. See, for instance, \citep{ZhangYang2017} and references therein.

To find $g(t)$, recall that according to \citep{EkstromTysk2012} for single-factor models that predict nonnegative short rates,  no second boundary condition is required for \eqref{PDE} if $r_t > 0, \ \forall t \ge 0$ in \eqref{OU1}, i.e., the boundary $r_t = 0$ is not attainable. Otherwise,  the PDE itself at this point serves as the boundary condition. In particular, as applied to our OU process, it reads
\begin{equation} \label{bc0}
 \left(\fp{F}{t} + \dfrac{1}{2}\sigma^2(t) \sop{F}{r} + \kappa(t) \theta(t) \fp{F}{r}\right)\Bigg|_{r=0} = 0.
\end{equation}
However, since  \eqref{OU1} is the OU process, it allows zero or even negative interest rates, i.e. in this model $r \in \mathbb{R}$. Therefore, the PDE \eqref{PDE} must be solved subject to the second boundary condition either at $r \to -\infty$, or at some artificial left boundary $r = r_{\min} < 0$.  It is not obvious, however,  how to setup this boundary condition at $r_{\min}$.

On the other hand, since the Hull-White model belongs to the class of affine models, \citep{andersen2010interest}, the solution of \eqref{PDE} can be represented in the form
\begin{equation} \label{affSol}
F(r,t,S) = A(t,S) e^{ B(t,S) r}.
\end{equation}
Substituting this expression into \eqref{PDE} and separating the terms proportional to $r$, we obtain two equations to determine $A(t,S), B(t,S)$
\begin{align} \label{equAff}
 \fp{B(t,S)}{t} &= B(t,S) \kappa (t) + 1, \\
 2 \fp{A(t,S)}{t} &= - A(t,S) B(t,S) \left[2 \theta (t) \kappa (t) +  B(t,S) \sigma (t)^2 \right]. \nonumber
\end{align}
To obey the terminal condition \eqref{termZCB}, the first PDE in \eqref{equAff} should be solved subject to the terminal condition $B(S,S) = 0$, and the second one - to  $A(S,S) = 1$. The solution reads
\begin{align} \label{ZCBsol}
B(t,S) &= e^{\int_0^t \kappa (x) \, dx} \int_S^t e^{-\int_0^x \kappa(q) \, dq} \, dx, \\
A(t,S) &= \exp\left[-\frac{1}{2} \int_S^t  B(x,S) \left( 2 \theta(x) \kappa(x) + B(x,S) \sigma^2(x) \right) dx \right]. \nonumber
\end{align}
It can be seen that $B(t,S) < 0$ if $t < S$. Therefore,  $F(r,t,S) \to 0$ when $r \to \infty$.

\section{Down-and-Out barrier option} \label{DOB}

Let us consider a Down-and-Out barrier Call option written on a ZCB as an underlying\footnote{Other types of underlying can also be handled within the proposed framework, eg., the LIBOR rate subject to the barrier conditions, etc., by the corresponding modification of \eqref{bcF}. Also, given the prices of barrier options on a ZCB, one can evaluate a continuous barrier caplet (flooret).}.  It is known that the price of the option $C(r,t)$ written on this bond under a risk-neutral measure solves the PDE, \citep{andersen2010interest}
\begin{equation} \label{PDEP}
\fp{C}{t} + \dfrac{1}{2}\sigma^2(t) \sop{C}{r} + \kappa(t) [\theta(t) - r] \fp{C}{r} = r C.
\end{equation}
This PDE should be solved subject to the terminal condition at the option maturity $T \le S$, and some boundary conditions provided to guarantee a uniqueness of the solution. The terminal condition reads
\begin{equation} \label{tc0}
C(r,T) = \left(F(r,T,S - K)\right)^+,
\end{equation}
\noindent where $K$ is the option strike.

By the contract definition, the lower barrier $L_{F}(t)$ is set on the Zero coupon bond (ZCB) price. In other words, at the barrier we have the following condition
\begin{equation} \label{ZCBBar}
C(r,t) = 0 \quad \mbox{if } F(r,t,S) = L_F(t).
\end{equation}
Since the ZCB price $F(r,t,S)$ is known in closed form in \eqref{affSol}, the above condition can be reformulated in the $r$ domain by solving the equation
\begin{equation} \label{Loft}
F(r,t,S) = A(t,S) e^{B(t,S) r} = L_F(t),
\end{equation}
\noindent with respect to $r$. This yields the following equivalent barrier $L(t)$ in the $r$ domain
\begin{equation} \label{Loft}
L(t) =  \frac{1}{B(t,S)} \log \left(\frac{L_F}{A(t,S)}\right) > 0,
\end{equation}
\noindent where it is assumed that $L_F(t) < A(t,S)$. Thus, the boundary condition to \eqref{PDEP} now reads
\begin{equation} \label{bcBar}
C(L(t),t) = 0.
\end{equation}

At the second boundary as $r \to \infty$ the ZCB price tends to zero, see \eqref{affSol}, and, therefore, the Call option price tends to zero. Thus, we set
\begin{equation} \label{bc01}
 C(r,t)\Big|_{r \to \infty}  = 0.
\end{equation}

Our goal now is to build a series of transformations to transform \eqref{PDEP} to the heat equation.

\subsection{Transformation to the heat equation} \label{trHeat}

To transform the PDE \eqref{PDE} to the heat equation we first rewrite \eqref{PDEP} in the form
\begin{equation} \label{PDEP2}
\fp{C}{t} = -\dfrac{1}{2}\sigma^2(t) \sop{C}{r} - [-\kappa(t)  r + \kappa(t)\theta(t)] \fp{C}{r} + r C.
\end{equation}
This equation belongs to the type of equations considered in \citep{Polyanin2002}, Section~3.8.7.  It is shown there that by transformation
\begin{equation} \label{transH}
C(r,t) = \exp[\alpha(t) r + \beta(t)] u(x,\tau), \qquad \tau = \phi(t), \qquad x = r \psi(t) + \xi(t),
\end{equation}
\eqref{PDEP2} can be reduced to the heat equation
\begin{equation} \label{Heat}
\fp{u}{\tau} = \sop{u}{x}.
\end{equation}
Here,
\begin{align} \label{coef}
\psi(t) &= C_1 \exp\left( \int_0^t \kappa(q) dq \right), \\
\phi(t) &= \frac{1}{2} \int_t^T  \sigma^2(q) \psi^2(q) d q + C_2, \nonumber \\
\alpha(t) &= \psi(t) \int_0^t \frac{1}{\psi(q)} dq + C_3 \psi(t), \nonumber \\
\beta(t) &= - \frac{1}{2} \int_0^t \alpha(q) \left[2 \kappa(q) \theta(q) + \sigma^2(q) \alpha(q) \right] dq + C_4, \nonumber \\
\xi(t) &= -\int_0^t  \left[ \kappa(q) \theta(q) + \sigma^2(q) \alpha(q) \right] \psi(q) dq + C_5, \nonumber
\end{align}
\noindent where $C_1,\ldots,C_5$ are some constants. In our case we can choose $C_1 = 1$, $C_2 = C_3 = C_4 = C_5  = 0$.

The \eqref{Heat} should be solved subject to the initial condition
\begin{align} \label{tc01}
u(x,0) &= \exp\left[ -\frac{\alpha(T)}{\psi(T)} (x - \xi(T)) - \beta(T) \right] \left(\bar{F}(x,T,S) - K\right)^+, \\
\bar{F}(x,T,S) &= A(T,S) \exp\left[  \frac{B(T,S)}{\psi(T)} (x - \xi(T)) \right], \nonumber
\end{align}
\noindent as it follows from \eqref{tc0}, \eqref{affSol} and \eqref{transH}.

The boundary conditions should be set at the new domain $x \in \Omega:  [y(\tau),\infty)$, where $y(\tau) = L(t(\tau)) \psi(t(\tau)) + \xi(t(\tau))$, and $t(\tau)$ is the inverse map $t \rightarrow \tau$. The latter  can be found explicitly by solving the second equation in \eqref{coef}
\begin{equation} \label{invMap}
\tau = \frac{1}{2} \int_t^T  \sigma^2(q) \psi^2(q) d q,
\end{equation}
\noindent with respect to t. Accordingly, the conditions at the boundaries of $\Omega$ can be obtained from \eqref{bc01}, \eqref{bcBar} and read
\begin{align} \label{bc}
u(x,\tau)\Big|_{x \to \infty}  &= 0, \qquad u(y(\tau),\tau) = 0.
\end{align}

\subsection{Solution of the barrier pricing problem} \label{GITmethod}

As applied to equities, the problem of pricing barrier options, where the underlying follows a time-dependent OU process, was considered in \citep{CarrItkin2020}. There the authors utilized and extended a method of generalized integral transform actively elaborated on by the Russian mathematical school to solve parabolic equations at the domain with moving boundaries, see eg., \citep{kartashov1999} and references therein. However, in \citep{CarrItkin2020,kartashov1999} those problems were solved at the domain $x \in [0,y(\tau)]$ while in this paper we have to deal with the infinite domain $\Omega$. For that kind of domains, the above method is not elaborated yet since there is a problem with constructing the inverse transform. Here we have managed to construct this solution (to the best of our knowledge, for the first time) and present it in Section~\ref{secSol}). Some examples where the inverse transform is not necessary can be found in \citep{kartashov1999, kartashov2001}.

But first we will attack this problem by using the method of heat potentials. This method is known in the theory of heat equation for a long time, see, eg., \citep{TS1963, Friedman1964, kartashov2001} and references therein. The first use of this method in mathematical finance is due to \citep{Lipton2002} for pricing path-dependent options with curvilinear barriers, and more recently in \citep{LiptonKu2018, LiptonPrado2020} (also see references therein).

Recall, that we must solve the heat equation in \eqref{Heat} with the initial condition in \eqref{tc01} and the boundary conditions in \eqref{bc}. Since this problem has an inhomogeneous initial condition, the method of heat potentials cannot be directly applied. Therefore, let us represent  $u(x,\tau)$ in the form
\begin{equation} \label{q}
u(x,\tau) = q(x,\tau) + \frac{1}{2\sqrt{\pi \tau}}  \int_{y(0)}^\infty u(x',0)   e^{-\frac{(x-x')^2}{4\tau}} dx'.
\end{equation}
Here the second term in the RHS is the solution of the heat equation in \eqref{Heat} at the infinite domain with the initial condition in \eqref{tc0}. But since $x \in \Omega$, we moved the left boundary from $-\infty$ to $y(0)$.

The function $q(x,\tau)$ solves the problem
\begin{align} \label{qP}
\fp{q(x,\tau)}{\tau} &= \sop{q(x,\tau)}{x}, \\
q(x,0) &= 0, \qquad y(0) < x < \infty, \nonumber \\
q(x,\tau)\Big|_{x \to \infty} &= 0, \qquad q(y(\tau),\tau) = \phi(\tau), \nonumber \\
\phi(\tau) &= - \frac{1}{2\sqrt{\pi \tau}}  \int_{y(0)}^\infty u(x',0)   e^{-\frac{(y(\tau)-x')^2}{4\tau}} dx'. \nonumber
\end{align}
This problem is similar to that in \eqref{Heat}, \eqref{tc0}, \eqref{bc}, but now with a homogeneous initial condition. Therefore, the solution can be found by the method of heat potentials.

Following the general idea of this method, we are looking for the solution of \eqref{qP} in the form of a generalized heat potential
\begin{equation} \label{poten}
q(x,\tau) = \frac{1}{4\sqrt{\pi}} \int_0^\tau \Psi(k) \frac{x-y(k)}{\sqrt{(\tau-k)^3}}e^{-\frac{(x-y(k))^2}{4(\tau-k)}} dk,
\end{equation}
\noindent  where $\Psi(k)$ is the heat potential density. It is easy to check that thus defined function $q(x,\tau)$ solves the first line of \eqref{qP}, and satisfies both the initial condition and the vanishing condition at $x \to \infty$. Also, from \eqref{poten} at the barrier $x = y(\tau)$ we must have, \citep{TS1963}
\begin{equation} \label{Abel2k}
2 \phi(\tau) = \Psi(\tau) + \frac{1}{2\sqrt{\pi}} \int_0^{\tau}  \Psi(k) \frac{y(\tau) - y(k)}{\sqrt{(\tau-k)^3}} e^{-\frac{(y(\tau)-y(k))^2}{4(\tau-k)}}  dk,
\end{equation}
\noindent since for $x = y(\tau)$ function $q(x,\tau)$ is discontinuous, and its limiting value at $x = y(\tau) + 0$ is equal to $\phi(\tau)$.

The \eqref{Abel2k} is a linear Volterra equations of the second kind, \citep{polyanin2008handbook}. Since $\phi(\tau)$ is a continuously differentiable function, \eqref{Abel2k} has a unique continuous solution for $\Psi(\tau)$. First, it can be efficiently solved by a standard approach,  i.e., the integral in the RHS is approximated using some quadrature rule with $N$ nodes in $k$, and the solution is obtained at $M$ nodes in $\tau$. This discretization gives rise to a linear equation with respect to the discrete vector of values of $\Psi(\tau)$ with the matrix being lower triangular. Therefore, the solution can be obtained by a simple elimination procedure with quadratic complexity. Alternatively, since the kernel is proportional to Gaussian, the discrete sum approximating the integral can be computed with linear complexity $O(N + M)$ using the Fast Gauss Transform, see eg., \citep{FGT2010}.  Then the final solution can be obtained as this is discussed in Section~\ref{numRes}.

Second, if in \eqref{invMap} $\tau(0)$ is small, we can approximate a curvilinear boundary $y(\tau)$ by a linear function
\begin{align} \label{app}
y(\tau) &= a + b \tau, \qquad a = y(0), \ b = \frac{y(\tau(0)) - y(0)}{y(\tau(0))}.
\end{align}
Then the integral kernel becomes a function of the difference $\tau - k$, and so the integrand is a convolutional function. Thus, \eqref{Abel2k} can be solved by using the Laplace or G- transforms. For instance, with allowance for \eqref{app} let us re-write \eqref{Abel2k} in the form,  \citep{kartashov2001}
\begin{align} \label{Abel2}
\phi_1(\tau) &= \Psi_1(\tau)   + \frac{b}{2\sqrt{\pi}} \int_0^\tau  \frac{\Psi_1(k)}{\sqrt{\tau-k}}  dk, \\
\phi_1(\tau) &= 2 \phi(\tau) e^{b^2 \tau/4}, \quad \Psi_1(k) = \Psi(k) e^{b^2 k/4}. \nonumber
\end{align}
This is the Abel integral equation of the second kind, {\citep{polyanin2008handbook}, which can be solved in closed form by using the Laplace transform. Since $\phi(0) = 0$, the solution reads
\begin{equation*}
\Psi_1(\tau) = {\cal F}(\tau) + \frac{b^2}{4} \int_0^\tau  e^{\frac{\tau-k}{4}} {\cal F}(k) dk,
\qquad {\cal F} = \phi_1(\tau) - \frac{b}{2\sqrt{\pi}} \int_0^\tau  \frac{\phi_1(k)}{\sqrt{\tau-k}} dk.
\end{equation*}
In case where the linear approximation is too crude, this expression can be used as a smart initial guess for the function $\Psi(\tau)$ which is needed by the iterative numerical method described in above.

Once \eqref{Abel2k} is solved and the function $\Psi(\tau)$ is found, the final solution reads
\begin{equation} \label{qSol}
u(x,\tau) = \frac{1}{4\sqrt{\pi}} \int_0^\tau  \Psi(k) \frac{x-y(k)}{\sqrt{(\tau-k)^3}}e^{-\frac{(x-y(k))^2}{4(\tau-k)}} dk + \frac{1}{2\sqrt{\pi \tau}}  \int_{y(0)}^\infty u(x',0)   e^{-\frac{(x-x')^2}{4\tau}} dx'.
\end{equation}
The second integral can be further simplified with allowance for the definition of $u(x,0)$ in \eqref{tc0}. This yields
\begin{align} \label{secI}
\frac{1}{2\sqrt{\pi\tau}} & \int_{y(0)}^\infty u(x',0)   e^{-\frac{(x-x')^2}{4\tau}} dx' \\
&= \frac{1}{2} e^{-\beta(T)} \Bigg[ e^{A_2 x + A_1} A(T,S)
\left(\mathrm{Erf}\left( \frac{x - y(0) + 2 \tau A_2}{2 \sqrt{\tau}}\right) -\mathrm{Erf}\left( \frac{x - K_1 + 2 \tau A_2}{2 \sqrt{\tau}}\right) \right) \nonumber \\
& - K e^{B_2 x + B_1}
\left(\mathrm{Erf}\left( \frac{x - y(0) + 2 \tau B_2}{2 \sqrt{\tau}}\right) -\mathrm{Erf}\left( \frac{x - K_1 + 2 \tau B_2}{2 \sqrt{\tau}}\right) \right) \Bigg], \nonumber \\
A_1 &= \frac{A_2}{\psi (T)} [\tau(B(T,S) - \alpha (T)) -\xi (T) \psi (T)], \qquad
A_2 = \frac{B(T,S) - \alpha(T)}{\psi(T)}, \nonumber \\
B_1 &= \frac{B_2}{\psi (T)} [-\tau \alpha(T) - \xi (T) \psi (T)], \qquad
B_2 = - \frac{\alpha (T)}{\psi (T)},  \nonumber \\
K_1 &= \max\Big\{ \xi(T) + \frac{\psi (T) }{B(T,S)} \log \left(\frac{K}{A(T,S)}\right), y(0)\Big\}.\nonumber
\end{align}
Also, by definition in \eqref{qP}, the function $\phi(\tau)$ can be represented in the same form just by substituting $x = y(\tau)$ and multiplying the result by -1.

\subsection{Second solution of the barrier pricing problem} \label{secSol}

In this Section we solve the same problem but using the method of generalized integral transform. This method was invented by the Russian mathematical school in the 20th century starting from A.V.~Luikov, and then by B.Ya.~Lyubov, E.M.~Kartashov, and some others, see a detailed survey in \citep{kartashov1999}. However, as mentioned in \citep{kartashov2001}, the solution for  a semi-infinite domain is not known yet, while some recommendations were given on how one can try to proceed.  Therefore, to the best of authors' knowledge, the solution presented in this Section is new and compliments the method of heat potentials presented in Section~\ref{GITmethod}.

We attack this problem by introducing the following integral transform
\begin{equation} \label{GITdef}
\baru(p,\tau) = \int_{y(\tau)}^\infty u(x,\tau) e^{-\sqrt{p}x} dx,
\end{equation}
where $p = a + i\omega$ is a complex number with $\operatorname{Re}(p) =\beta > 0$, and $-\pi/4 < \arg(\sqrt{p}) < \pi/4$.
Let us multiply both parts of \eqref{Heat} by $e^{-x\sqrt{p}}$ and then integrate on $x$ from $y(\tau)$ to $\infty$:
\begin{align} \label{tr2}
\frac{\partial}{\partial \tau} &\int_{y(\tau)}^\infty u(x,\tau) e^{-\sqrt{p}x} dx + u(y(\tau),\tau) e^{-\sqrt{p}y(\tau)} y'(\tau) = \\
&= \frac{\partial u}{\partial x} e^{-\sqrt{p}x}\Big|_{x = y(\tau)}^{x \rightarrow+\infty} + \sqrt{p} u(x,\tau)  e^{-\sqrt{p}x} \Big|_{x = y(\tau)}^{x \rightarrow+\infty} +  p \int_{y(\tau)}^\infty u(x,\tau) e^{-\sqrt{p}x} dx. \nonumber
\end{align}

By taking into account the boundary conditions in \eqref{bc}, \eqref{tr2} can be represented as the following Cauchy problem
\begin{align} \label{barUeq}
\frac{d\baru}{d \tau} - p \bar{u} &= \Upsilon(\tau) e^{-\sqrt{p}y(\tau)}, \\
\bar{u}(p, 0) &= \int_{y(0)}^\infty u(x,0) e^{-\sqrt{p}x} dx, \quad \Upsilon(\tau) = -\frac{\partial u(x,\tau)}{\partial x} \big|_{x = y(\tau)}, \nonumber
\end{align}
\noindent where $u(x,0)$ us given in \eqref{tc01}. The solution of this problem reads
\begin{equation}  \label{barU_explicit}
\baru(p,\tau) = e^{p\tau} \int_0^\tau \Upsilon(s) e^{-ps} e^{-\sqrt{p}y(s)}ds
+ e^{p\tau} \int_{y(0)}^\infty u(z,0) e^{-\sqrt{p}z} dz.
\end{equation}
By analogy with \citep{CarrItkin2020},  the function $\Upsilon(\tau)$ solves the Fredholm equation of the first type
\begin{equation} \label{FredholmPsi}
\int_0^{\infty} \Upsilon(\tau) e^{-p\tau - \sqrt{p} \tau} d\tau =F(p),
\end{equation}
\noindent with
\begin{align} \label{FforFredholm}
F(p) = - e^{\frac{\alpha(T) \xi(T)}{\psi(T)} - \beta(T)}
&\left[
A(T,S) e^{\frac{B(T,S)\xi(T)}{\psi(T)}} \frac{e^{-f_1(p,T) y_0} -e^{-f_1(p,T) K_1}}{f_1(p,T)} -K \frac{e^{-f_2(p,T) y_0} -e^{-f_2(p,T)  K_1}}{f_2(p,T)}
\right]
\\
f_1(p, T) &= \sqrt{p} + \frac{\alpha(T)}{\psi(T)} - \frac{\beta(T)}{\psi(T)}, \quad f_2(p, T) = \sqrt{p} + \frac{\alpha(T)}{\psi(T)}, \nonumber
\end{align}
\noindent where $K_1$ is defined in \eqref{secI}.

To construct the inverse transform, recall that  the solution of the heat equation $\mathcal{L}u(x, \tau ) = 0$, $\mathcal{L} = \partial / \partial \tau - \partial^2 / \partial x^2$ in the half-plane domain $x \in (0, \infty)$ can be expressed via the Fourier sine integral, \citep{cannon1984one}
\[ \int_0^\infty \alpha(\xi) e^{-\xi^2 \tau} \sin(\xi x) d\xi. \]
Therefore, by analogy we look for the inverse transform of $\bar{u}$ (or for the solution of \eqref{barUeq} in terms of $u$) to be an oscillatory integral of the form
\begin{equation} \label{invTr}
u(x,\tau) = \int_0^\infty \alpha(\xi, \tau) \sin[\xi(x- y(\tau))] d\xi,
\end{equation}
\noindent where $\alpha(\xi, \tau)$ is some function to be determined. Note, that this definition automatically respects the vanishing boundary conditions for $u(x,\tau)$. We assume that this integral converges absolutely and uniformly $\forall x \in [y(\tau), \infty)$ for any $\tau > 0$.

Applying \eqref{GITdef} to both parts of \eqref{invTr} and integrating yields
\begin{align}
\bar{u}(p,\tau) &= \int_{y(\tau)}^\infty e^{-\sqrt{p}x} \int_0^\infty \alpha(\xi, \tau) \sin(\xi(x- y(\tau))) d\xi dx
= e^{-\sqrt{p} y(\tau)} \int_0^\infty   \alpha(\xi, \tau)  \frac{ \xi  d\xi}{\xi^2 + p}.
\end{align}

Replacing $\bar{u}(p,\tau)$ with the solution found in \eqref{barU_explicit}, we obtain
\begin{equation} \label{InvEq}
\int_0^\infty   \alpha(\xi, \tau) \frac{ \xi  d\xi}{\xi^2 + p} = \int_0^\tau \Upsilon(s) e^{p(\tau-s)} e^{\sqrt{p}(y(\tau) - y(s))}ds + e^{p\tau}\int_{y(0)}^\infty u(z,0) e^{\sqrt{p}(y(\tau) - z)} dz.
\end{equation}
Now, similar to inverse operator methods, like the inverse Laplace transform, we need an analytic continuation of the transform parameter $p$ into the  complex plane. Let us integrate both sides of \eqref{InvEq} on $p$ along the so-called keyhole contour presented in Fig.~\ref{contour}.
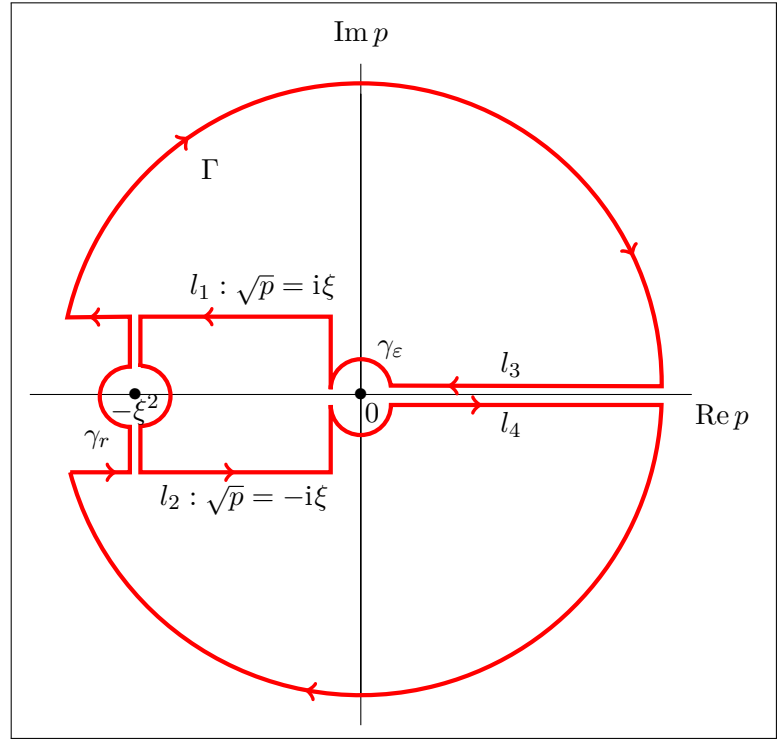
\begin{figure}[!ht]
\hspace{-0.5in}
\captionsetup{width=0.8\linewidth}	
\begin{center}
\fbox{
\begin{tikzpicture}
% Configurable parameters
\def\bigradius{4}
\def\axisradius{4}
\def\gammaradius{0}
\def\omegaradius{0}
\def\littleradius{0.4}
\def\xii{15}
\def\ang{10}
\def\sh{0.07}
\def\pole{3}

% Axes
\draw (-1.1*\axisradius, 0) -- (1.1*\axisradius,0)
      (\omegaradius, -1.1*\axisradius) -- (\omegaradius, 1.1*\axisradius)
      (\gammaradius, -\bigradius) -- (\gammaradius, \bigradius);

\draw[red, ultra thick, decoration={ markings,
        mark=at position 0.1 with {\arrow{<}}
        ,mark=at position 0.3 with {\arrow{<}}
        ,mark=at position 0.4 with {\arrow{<}}
        ,mark=at position 0.5 with {\arrow{<}}
        ,mark=at position 0.6 with {\arrow{<}}
        ,mark=at position 0.7 with {\arrow{<}}
        ,mark=at position 0.85 with {\arrow{<}},
        ,mark=at position 0.92 with {\arrow{<}}
        ,mark=at position 0.99 with {\arrow{<}}
	},
    postaction={decorate}]
    let
        \n1 = {asin(\sh/\littleradius)},
        \n2 = {\bigradius*sin(\xii)},
        \n3 = {\littleradius*sin(\ang)},
        \n4 = {\bigradius*cos(\xii)},
        \n5 = {\bigradius*sin(\xii)},
        \n6 = {\bigradius*sin(2)},
        \n7 = {\bigradius*cos(2)}
    in
		 (-\n4,-\n5) arc(-180+\xii:-2:\bigradius) -- (\littleradius, -\n6) -- (\littleradius, -\n6) arc(0:-180:\littleradius) -- (-\littleradius, -\n2)
        --  (-\pole+\sh, -\n5)  -- (-\pole+\sh, -\n3-0.35) -- (-\pole+\sh, -\n3-0.35) arc(-90:90:\littleradius)
        -- (-\pole+\sh, \n3+0.3) -- (-\pole+\sh, \n5) -- (-\littleradius, \n5) -- (-\littleradius, \n3)
        -- (-\littleradius, \n3) arc(180:7:\littleradius) --  (\bigradius, \n6-0.03) -- (\bigradius, \n6)  arc(0:180-\ang-2.8:\bigradius)
        --  (-\pole-\sh, \n5) --  (-\pole-\sh, \n3+0.3)
        --  (-\pole-\sh, \n3+0.3) arc(90:270:\littleradius) -- (-\pole-\sh, -\n5)  -- (-\n4,-\n5);

%% The labels
\node at (1.2*\axisradius,-0.3){$\operatorname{Re} p$};
\node at (0,1.2*\axisradius) {$\operatorname{Im} p$};
\node at (0.15,-0.25){$0$};
\node at (0.4,0.6) {$\gamma_{\varepsilon}$};
\node at (-\pole-0.5,-0.6) {$\gamma_{r}$};
\node at (0.,0) {$\bullet$};
\node at (-\pole,0) {$\bullet$};
\node at (-\pole,-0.2) {$-\xi^2$};
\node at (-2.,3.) {$\Gamma$};
%\node at (-1.1,1.4) {$l_1 = \iu \xi$};
%\node at (-1.555,-1.4) {$l_2 = - \iu \xi$};
\node at (-1.3,1.4) {$l_1 : \sqrt{p}= \iu \xi$};
\node at (-1.55,-1.4) {$l_2 :\sqrt{p} = - \iu \xi$};
\node at (\bigradius/2, 0.4) {$l_3$};
\node at (\bigradius/2, -0.4) {$l_4$};
\end{tikzpicture}
}
\end{center}
\caption{Contour of integration of \eqref{InvEq} in a complex plane of $p$.}
\label{contour}
\end{figure}
 In more detail, this contour can be described as follows.
 It starts with a big symmetric arc $\Gamma$ around the origin with the radius $R$;
 extending to two horizontal line segments $l_3, l_4$ (a cut around the line $\operatorname{Im} p = 0, \operatorname{Re} p > 0$);
  connecting to two small semi-circles $\gamma_\varepsilon$ around the origin with the radius $\varepsilon \ll 1$;
   then extending to two vertical line segments up to points  $\operatorname{Im} p = \pm \xi$;
    then again two horizontal parallel line segments  $l_1, l_2$ at $\operatorname{Im} p = \pm \xi$, which end points are connected to the arc $\Gamma$ with a cut at $\operatorname{Im} p = -\xi^2$ (it consists of two vertical line segments and two semi-circles $\gamma_r$ with the radius $\varepsilon$), such that the whole contour is continuous.

 Using a standard technique, we take a limit $\varepsilon \to 0, R \to \infty$, so in this limit the contour takes the form as depicted in Fig,~\ref{Fig2}. It has a horizontal cut along the positive real line with point $p=0$ excluded from the area inside the contour; another vertical cut at $\operatorname{Re} p = -\xi^2$ with the point $p = -\xi^2$ lying inside the contour; and a branch cut $l_1, l_2$ of the multivalued function $\sqrt{p}$ at $p = -\xi^2$. Also, in this limit $l_7 \to 0, l_8 \to 0$, but in Fig.~\ref{Fig2} we left them as it is for a better readability.

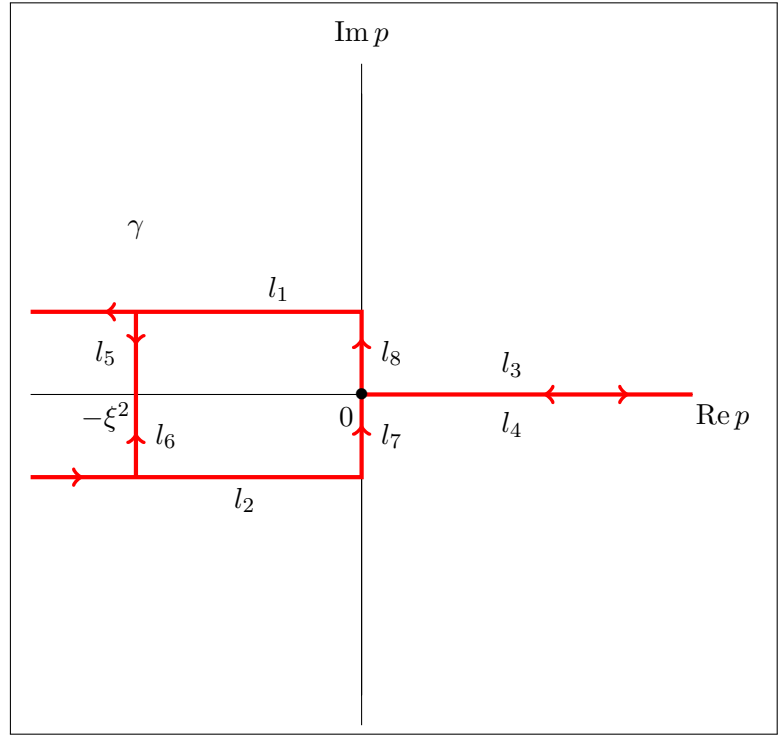
\begin{figure}[!ht]
\hspace{-0.5in}
\captionsetup{width=0.8\linewidth}	
\begin{center}
\fbox{
\begin{tikzpicture}
% Configurable parameters
\def\bigradius{4}
\def\axisradius{4}
\def\gammaradius{0}
\def\omegaradius{0}
\def\littleradius{0.4}
\def\xii{15}
\def\ang{10}
\def\sh{0.07}
\def\pole{3}
\def\vert{1.1}

% Axes
\draw (-1.1*\axisradius, 0) -- (1.1*\axisradius,0)
      (\omegaradius, -1.1*\axisradius) -- (\omegaradius, 1.1*\axisradius)
      (\gammaradius, -\bigradius) -- (\gammaradius, \bigradius);

\draw[red, ultra thick, decoration={ markings,
        mark=at position 0.1 with {\arrow{<}}
        ,mark=at position 0.2 with {\arrow{>}}
        ,mark=at position 0.52 with {\arrow{>}}
        ,mark=at position 0.9 with {\arrow{>}}
	},
    postaction={decorate}]
    (1.1*\bigradius, 0) -- (0,0) -- (0,\vert) -- (-1.1*\bigradius,\vert);

\draw[red, ultra thick, decoration={ markings,
        mark=at position 0.1 with {\arrow{<}}
        ,mark=at position 0.9 with {\arrow{<}}
	},
    postaction={decorate}]
    (0,0) -- (0,-\vert) -- (-1.1*\bigradius,-\vert);

\draw[red, ultra thick, decoration={ markings,
        mark=at position 0.2 with {\arrow{>}}
        ,mark=at position 0.8 with {\arrow{<}}
	},
    postaction={decorate}]
    (-\pole,\vert) -- (-\pole,-\vert);

%% The labels
\node at (1.2*\axisradius,-0.3){$\operatorname{Re} p$};
\node at (0,1.2*\axisradius) {$\operatorname{Im} p$};
\node at (-0.2,-0.3){$0$};
\node at (0.,0) {$\bullet$};
\node at (-\pole-0.4,-0.3) {$-\xi^2$};
\node at (-1.1,1.4) {$l_1$};
\node at (-1.555,-1.4) {$l_2$};
\node at (\bigradius/2, 0.4) {$l_3$};
\node at (\bigradius/2, -0.4) {$l_4$};
%\node at (0.3, \vert) {$\xi$};
%\node at (0.3, -\vert) {$-\xi$};
\node at (-\pole-0.4,0.5*\vert) {$l_5$};
\node at (-\pole+0.4,-0.5*\vert) {$l_6$};
\node at (0.4,-0.5*\vert) {$l_7$};
\node at (0.4,0.5*\vert) {$l_8$};
\node at (-\pole, 2*\vert) {$\gamma$};
\end{tikzpicture}
}
\end{center}
\caption{Contour of integration $\gamma$ of \eqref{InvEq} in a complex plane of $p$ at $\varepsilon \to 0, R \to \infty$.}
\label{Fig2}
\end{figure}

Now we are ready to compute the integrals in \eqref{InvEq}. That one in the LHS is regular everywhere inside this contour except a single pole $p = -\xi^2$. By the residue theorem, we obtain
\begin{equation} \label{Int1}
\oint_{\gamma}  \left( \int_0^\infty  \alpha(\xi, \tau) \frac{\xi d\xi}{\xi^2 + p} \right)dp = -2\pi i \int_0^\infty \xi \alpha(\xi, \tau) d\xi.
\end{equation}

The integral in the RHS doesn't have any singularity inside the contour $\gamma$, however, it has several cuts. As can be easily checked, the integrals along the segments $l_3$ and $l_4$ cancel out, as well as those along $l_7$ and $l_8$, and those along $l_5$ and $l_6$.  The integral along the contour $\Gamma$ tends to zero if $R \to \infty $ due to Jordan's lemma. Hence, the only remaining integrals are those along the horizontal semi-infinite lines $l_1$ and $l_2$. They read
\begin{align} \label{Int2}
\int_{l_1} & \left( \int_0^\tau \Upsilon(s) e^{p(\tau-s)} e^{\sqrt{p}(y(\tau) - y(s))}ds
+ e^{p\tau} \int_{y(0)}^\infty u(z,0) e^{\sqrt{p}(y(\tau) - z)} dz \right)dp \\
&= -2\int_0^\infty \xi\left( \int_0^\tau \Upsilon(s) e^{-\xi^2(\tau-s)} e^{i\xi(y(\tau) - y(s))} ds
+ e^{-\xi^2\tau} \int_{y(0)}^\infty u(z,0) e^{i\xi(y(\tau) - z)} dz \right)d\xi, \nonumber \\
\int_{l_2}& \left( \int_0^\tau \Upsilon(s) e^{p(\tau-s)} e^{\sqrt{p}(y(\tau) - y(s))} ds
+ e^{p\tau} \int_{y(0)}^\infty u(z,0) e^{\sqrt{p}(y(\tau) - z)} dz \right) dp \nonumber \\
&=2\int_0^\infty \xi\left( \int_0^\tau \Upsilon(s) e^{-\xi^2(\tau-s)} e^{-i\xi(y(\tau) - y(s))}ds
+ e^{-\xi^2\tau} \int_{y(0)}^\infty u(z,0) e^{-i\xi(y(\tau) - z)} dz \right)d\xi. \nonumber
\end{align}
The RHS of \eqref{InvEq} is equal to a sum of these integrals
\begin{equation*}
4i \int_0^{\infty} \xi \left[\int_0^\tau \Upsilon(s) e^{-\xi^2(\tau-s)} \sinh\left( i\xi(y(\tau) - y(s)) \right)ds
+ e^{-\xi^2\tau} \int_{y(0)}^\infty u(z,0) \sin\left(\xi(y(\tau) - z\right) dz \right] d\xi.
\end{equation*}
Equating the LHS and RHS provides an explicit representation of $\alpha(\xi, \tau)$
\begin{equation} \label{alpha_explicit}
\alpha(\xi, \tau) = -\frac{2}{\pi}\left[\int_0^\tau \Upsilon(s) e^{-\xi^2(\tau-s)} \sin\left(\xi(y(\tau) - y(s)) \right)ds
+ e^{-\xi^2\tau} \int_{y(0)}^\infty u(z,0) \sin\left(\xi(y(\tau) - z\right) dz \right]
\end{equation}
Finally, we substitute \eqref{alpha_explicit} into \eqref{InvEq} and take into account the identity, \citep{GR2007}
\begin{equation*}
\int_0^\infty e^{-\beta x^2} \sin\left(ax\right) \sin\left(bx\right) dx =
\frac{1}{4} \sqrt{\frac{\pi}{\beta}} \left(  e^{-\frac{(a-b)^2}{4\beta}} - e^{-\frac{(a+b)^2}{4\beta}} \right)
,\quad \beta > 0,
\end{equation*}
\noindent which yields
\begin{align} \label{u_explicit}
u(x,\tau) &= \frac{1}{2\sqrt{\pi}} \int_0^\tau \frac{\Upsilon(s)}{\sqrt{\tau - s}}
\left( e^{-\frac{(x - y(s))^2}{4(\tau - s)}} - e^{-\frac{(x - 2y(\tau) + y(s))^2}{4(\tau - s)}} \right) ds \\
&+ \frac{1}{2\sqrt{\pi \tau}}\int_{y(0)} ^{\infty} u(z,0) \left( e^{-\frac{(x - z)^2}{4\tau}} - e^{-\frac{(x - 2y(\tau) + z)^2}{4\tau}}
\right) dz. \nonumber
\end{align}
Thus, we obtained another representation of the solution which reads a bit different from that in \eqref{qSol}, despite the general ansatz looks similar. Indeed, the solution is a sum of two integrals: one with respect to time $\tau$, and the other one with respect to $x$ with the integrand being a product of the initial condition with a Gaussian weight (it can be computed explicitly via Erf functions). The difference can be attributed to the different definitions of function $\Upsilon(k)$, as in \eqref{poten} it is the heat potential density, while in \eqref{barUeq} this is the gradient of the solution at $x = y(\tau)$. The first one is determined by the solution of the Volterra equation of the second  kind \eqref{Abel2k}, and the  second one - by the solution of the Fredholm equation of the first kind in \eqref{FredholmPsi}. However, the latter can either be transformed to the Volterra equation of the second kind. For doing that, one needs to differentiate \eqref{u_explicit} on $x$, and then let $x = y(\tau)$. This yields
\begin{align} \label{Fin}
 \Upsilon(\tau) &= \frac{1}{2\sqrt{\pi}} \int_0^\tau \Upsilon(s) \frac{y(\tau) - y(s)}{(\tau - s)^{3/2}} e^{-\frac{(y(\tau) - y(s))^2}{4(\tau - s)}}  ds
+ \frac{1}{2\sqrt{\pi \tau^3}}\int_{y(0)} ^{\infty} u(z,0) (y(\tau) - z) e^{-\frac{(z-y(\tau))^2}{4\tau}} dz.
\end{align}

\section{Double barrier options} \label{DBO}

Similar to Section~\ref{GITmethod},\ref{secSol}, both the HP and GIT methods can be used to price double barrier options with the lower barrier $L_F(t)$ and the upper barrier $H_F(t)$. Here we present the solution obtained with the HP method, while the GIT approach will be published elsewhere.

For pricing double barrier options we have the following problem to solve
\begin{align} \label{uDB}
\fp{u(x,\tau)}{\tau} &= \sop{u(x,\tau)}{x}, \\
u(x,\tau=0) &= u(x,0), \qquad y(0) < x < z(0), \nonumber \\
u(y(\tau),\tau) &= u(z(\tau), \tau) = 0, \nonumber
\end{align}
\noindent where $z(\tau) = H(t(\tau)) \psi(t) + \xi(t)$ is the moving upper boundary, $H(t(\tau)$ is defined in \eqref{Loft} by replacing $L_F$ with $H_F$, and $u(x,0)$ is defined in \eqref{tc01}.

Similar to \eqref{q} we represent the solution in the form
\begin{equation} \label{q1}
u(x,\tau) = q(x,\tau) + \frac{1}{2\sqrt{\pi \tau}}  \int_{y(0)}^{z(0)} u(x',0)   e^{-\frac{(x-x')^2}{4\tau}} dx',
\end{equation}
\noindent so the function $q(x,\tau)$ solves a problem with the homogeneous initial condition
\begin{align} \label{qDB}
\fp{q(x,\tau)}{\tau} &= \sop{q(x,\tau)}{x}, \\
q(x,0) &= 0, \qquad y(0) < x < z(0), \nonumber \\
q(y(\tau),\tau) &= - \phi_2(\tau), \qquad q(z(\tau), \tau) = -\psi_2(\tau), \nonumber \\
\phi_2(\tau) =  - \frac{1}{2\sqrt{\pi \tau}} & \int_{y(0)}^{z(0)} u(x',0)   e^{-\frac{(y(\tau)-x')^2}{4\tau}} dx', \qquad
\psi_2(\tau) = - \frac{1}{2\sqrt{\pi \tau}}  \int_{y(0)}^{z(0)} u(x',0)   e^{-\frac{(z(\tau)-x')^2}{4\tau}} dx'. \nonumber
\end{align}

Again, we are looking for the solution of \eqref{qDB} in the form of a generalized heat potential
\begin{equation} \label{poten1}
q(x,\tau) = \frac{1}{4\sqrt{\pi}} \int_0^\tau  \frac{1}{\sqrt{(\tau-k)^3}} \left( (x-y(k)) \Psi(k) e^{-\frac{(x-y(k))^2}{4(\tau-k)}}
+ (x-z(k))\Phi(k) e^{-\frac{(x-z(k))^2}{4(\tau-k)}} \right) dk,
\end{equation}
\noindent  where $\Psi(k), \Phi(k)$ are the heat potential densities. They solve a system of two Volterra equations of the second kind
\begin{align} \label{Abel2k1}
2\phi_2(\tau) &= \Psi(\tau) + \frac{1}{2\sqrt{\pi}} \int_0^{\tau}   \left( \Psi(k) \frac{y(\tau)-y(k)}{(\tau-k)^{3/2} } e^{-\frac{(y(\tau)-y(k))^2}{4(\tau-k)}}
+\Phi(k) \frac{y(\tau)-z(k)}{(\tau-k)^{3/2} } e^{-\frac{(y(\tau)-z(k))^2}{4(\tau-k)}} \right) dk, \\
2\psi_2(\tau) &= - \Phi(\tau) + \frac{1}{2\sqrt{\pi}} \int_0^{\tau} \left( \Psi(k) \frac{z(\tau)-y(k)}{(\tau-k)^{3/2} } e^{-\frac{(z(\tau)-y(k))^2}{4(\tau-k)}}
+ \Phi(k) \frac{z(\tau)-z(k)}{(\tau-k)^{3/2} }e^{-\frac{(z(\tau)-z(k))^2}{4(\tau-k)}} \right) dk, \nonumber
\end{align}
\noindent where functions $\psi_2(\tau), \phi_2(\tau)$ can be expressed in closed form. This system, can be solved either by using quadratures (which results in solving a linear system with lower triangular matrix with the quadratic complexity but with no iterations), or by the Variational Iteration Method (VIM), see \citep{Wazwaz2011} with a linear complexity by using the Fast Gauss transform. Once this is done, the solution of our double barrier problem is found.

\section{Numerical example} \label{numRes}

To test performance and accuracy of our method we run a test where the explicit form of parameters $\kappa(t), \theta(t), \sigma(t)$ is chosen as
\begin{equation} \label{ex}
\kappa(t) = \kappa_0, \qquad \theta(t) = \theta_0 e^{-\theta_k t}, \qquad \sigma(t) = \sigma_0 e^{-\sigma_k t},
\end{equation}
\noindent with $\kappa_0, \theta_0, \sigma_0, \theta_k, \sigma_k$ being constants. With these definitions all  functions in \eqref{coef} can be found in closed form.

We approach pricing of Down-and-Out barrier Call option written on a ZCB in two ways. First, we solve the PDE in \eqref{PDEP}  by using a finite-difference scheme of the second order in space and time. We use the Crank-Nicolson scheme with few first Rannacher steps on a non-uniform grid compressed close to the barrier level at $t=0$, in more detail, see \citep{ItkinBook}. Second, we use the method of heat potentials (HP) and solve the same problem as this is described in Section~\ref{GITmethod}. To solve the Volterra equation in \eqref{Abel2k} we approximate the kernel on a rectangular grid $M \times M$, and the integral using the trapezoidal rule, which gives rise to the following system of linear equations
\begin{equation} \label{matEq}
\|2 \phi \| = (I + P) \| \Psi \|.
\end{equation}
Here $\| \Psi \|$ is the vector of discrete values of $\Psi(\tau), \ \tau \in [0, \tau(0)]$ on a grid with $M$ nodes, $\| \phi \|$ is a similar vector of $\phi(\tau)$, $I$ is the unit $M\times M$ matrix, and $P$ is the $M\times M$ matrix  of the kernel values on the same grid. Due to the specific structure of the Gaussian kernel, matrix $P$ is lower triangular. Therefore, solution of \eqref{matEq} can be done with complexity $O(M^2)$.

An important point here is that the kernel (and so the matrix $P$) doesn't depend of strikes $K$, but only the function $\phi(\tau)$. Therefore, \eqref{matEq} can be solved simultaneously for all strikes by inverting the matrix $I + P$ with the complexity $O(M^2)$, and then multiplying it by vectors $\| \phi \|_k, \ k=1,\ldots,\bar{k}$,  $\bar{k}$ is the total number of strikes. Hence, the total complexity is $O(\bar{k}M^2)$.

We emphasize, that this algorithm of solving the Volterra equation \eqref{Abel2k} doesn't require iterations, as that which makes use of the FGT. To compare both algorithms, note that complexity of the matrix algorithm is $O(M^2)$ versus $O(2 n M)$ - the complexity of the iterative algorithm which requires $n$ iterations to converge. Therefore, if $M$ is relatively small, eg., $M = 20$, and the number of iterations is near 10, both algorithms are of the same complexity.

For this test parameters of the model are presented in Table~\ref{tab1}.

\begin{table}[!htb]
\begin{center}
\begin{tabular}{|c|c|c|c|c|c|c|c|c|c|}
\hline
$r_0$ & $\kappa_0$ & $\theta_0$ & $\sigma_0$ & $\theta_k$ & $\sigma_k$ & $L_F$   &$S$\\
\hline
0.07 & 1.0 & 0.08 & 0.2 & 0.3 & 0.2 & 0.8 & 7  \\
\hline
\end{tabular}
\caption{Parameters of the test.}
\label{tab1}
\end{center}
\end{table}

We run the test for a set of maturities $T \in [1/12, 0.3,0.5,1]$ and strikes $K \in [0.06, 0.08, 0.1, 0.15, 0.2, 0.3]$. The Down-and-Out barrier Call option prices computed in such an experiment are presented in Fig.~\ref{figPrice}.
\noindent\begin{figure}[!htb]
\begin{minipage}{0.4\textwidth}
\centering
\captionsetup{width=\textwidth}
\vspace{-0.3in}
\includegraphics[totalheight=2.6in]{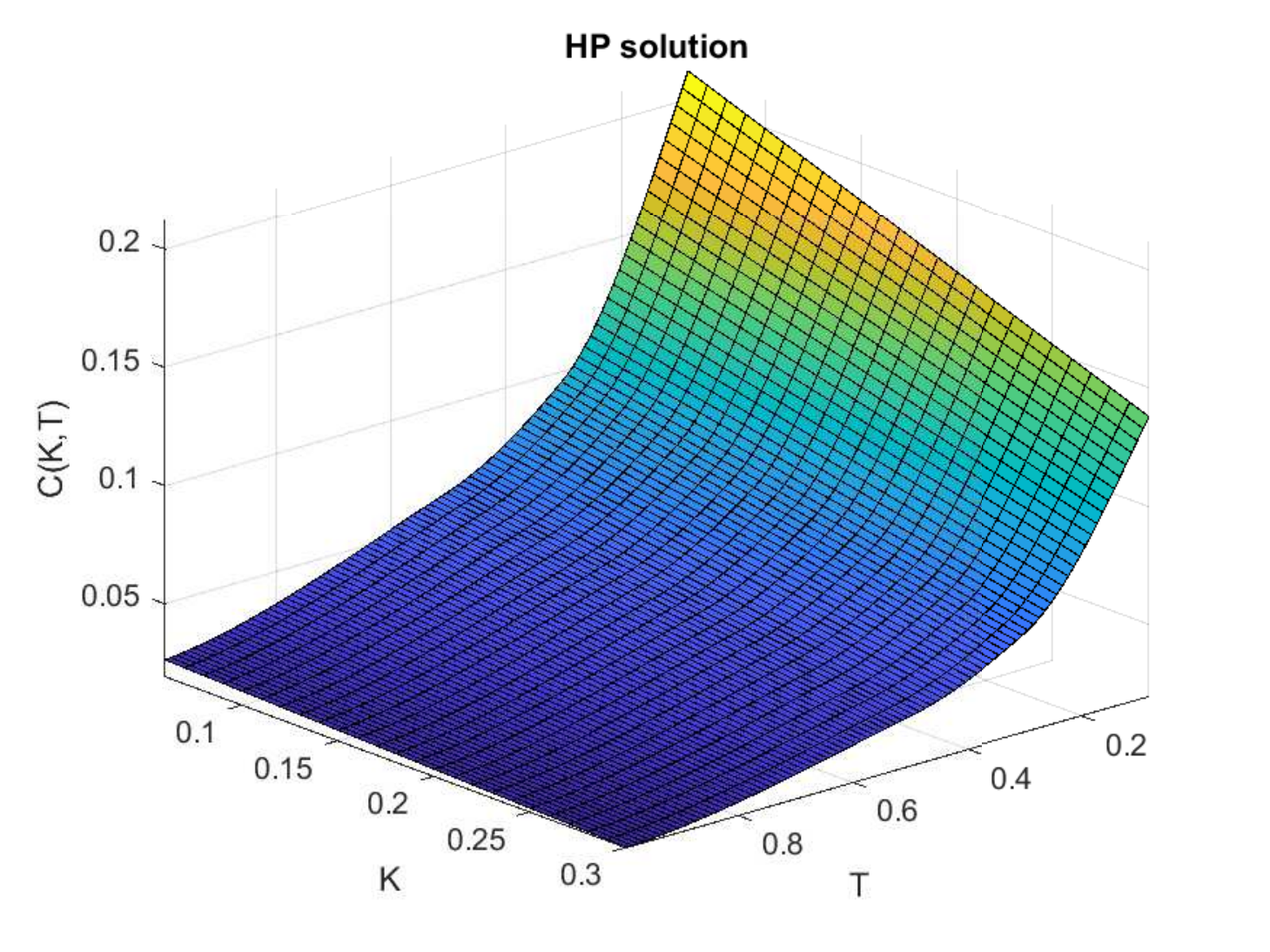}
\caption{Down-and-Out barrier Call option price computed by using the HP method.}
\label{figPrice}
\end{minipage}
\hspace{0.1\textwidth}
\begin{minipage}{0.4\textwidth}
\centering
\captionsetup{width=\textwidth}
\includegraphics[totalheight=2.6in]{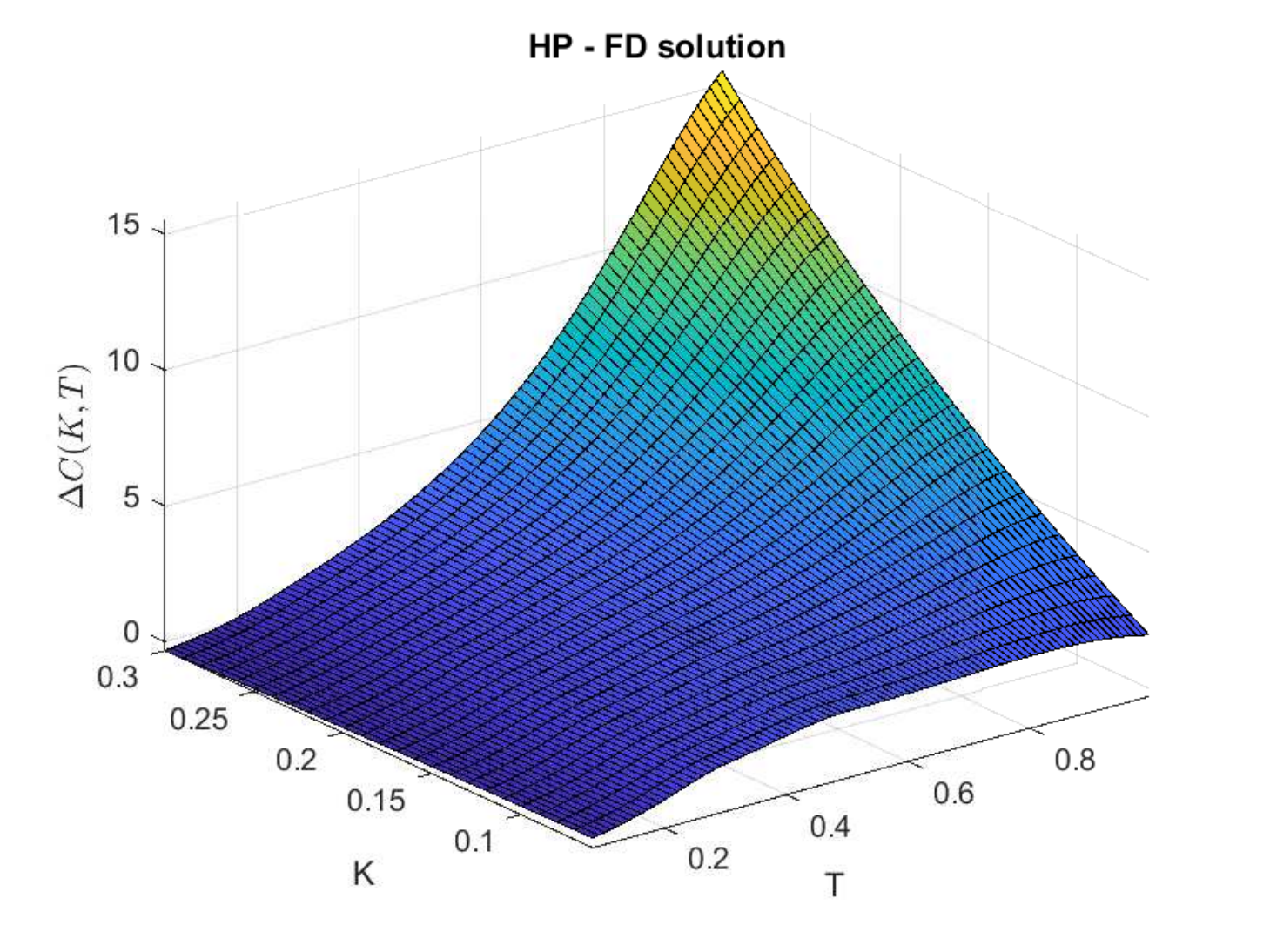}
\caption{The relative difference in \% of the Down-and-Out barrier Call option prices computed by using  the HP method and the FD solver with 201 nodes in space $r$ and time $t$.}
\label{error}
\end{minipage}
\end{figure}

In Fig.~\ref{error} the relative differences (in percents) between these prices obtained by using the HP method and the FD solver  are presented as a function of the option strike $K$ and maturity $T$. Here to provide a comparable accuracy we run the FD solver with 200 nodes in space $r$ and 201 steps in time $t$. Otherwise the quality of the FD solution is poor.

A large difference about 15\% at simultaneously high maturities and strikes is due to the very low computed price of the option, which is 1.62 cents for the FD method, and 1.92 cents for the HP method, respectively. Otherwise, at small $T$ and $K$ the difference is about 0.1-2\%. Also, since in this test we have chosen a fixed barrier in the ZCB price space, the corresponding barrier in the $r$ space is moving down. Therefore, the computed option price could vanish for strikes close to the barrier at short maturities, but contrary have some positive value for the same strikes at longer maturities.

Obviously, since the FD solver needs a high number of nodes in both space and time to achieve a reasonable accuracy, the cost for this is speed. Suppose that this solver uses $N_r$ nodes in space $r$, and $M_t$ nodes in time $t$. Then the total complexity of solving the forward PDE to simultaneously get option prices for all given strikes $K_i, \ i=1,\ldots,\bar{k}$ and maturities $T_j, \ j=1,\ldots,\bar{m}$ is $O(N_r \times M_t)$. This should be compared with the complexity of the HP method which is $O(\bar{k}\bar{m} M^2)$. Since, as we saw, $N_r = M_t = 200, \ \bar{k} = 6, \ \bar{m} = 4, \ M = 20$, the HP method should be four times faster than the FD method. In reality, our test with 24 points in the $K\times T$ space shows that the elapsed time of the HP method is 20 mls, while for the FD method this is 146 mls which is 7 times slower. Decreasing the size of the FD grid to $100\times 100$ nodes also decreases the elapsed time to 21 mls, but by the cost of increasing a relative difference up to $\pm 15$\% for a wide range of maturities and strikes. Thus, overall, the method of HP demonstrates, at least same performance as the forward FD solver.

\begin{figure}[!htb]
\centering
\captionsetup{width=\textwidth}
\vspace{-0.3in}
\includegraphics[totalheight=2.7in]{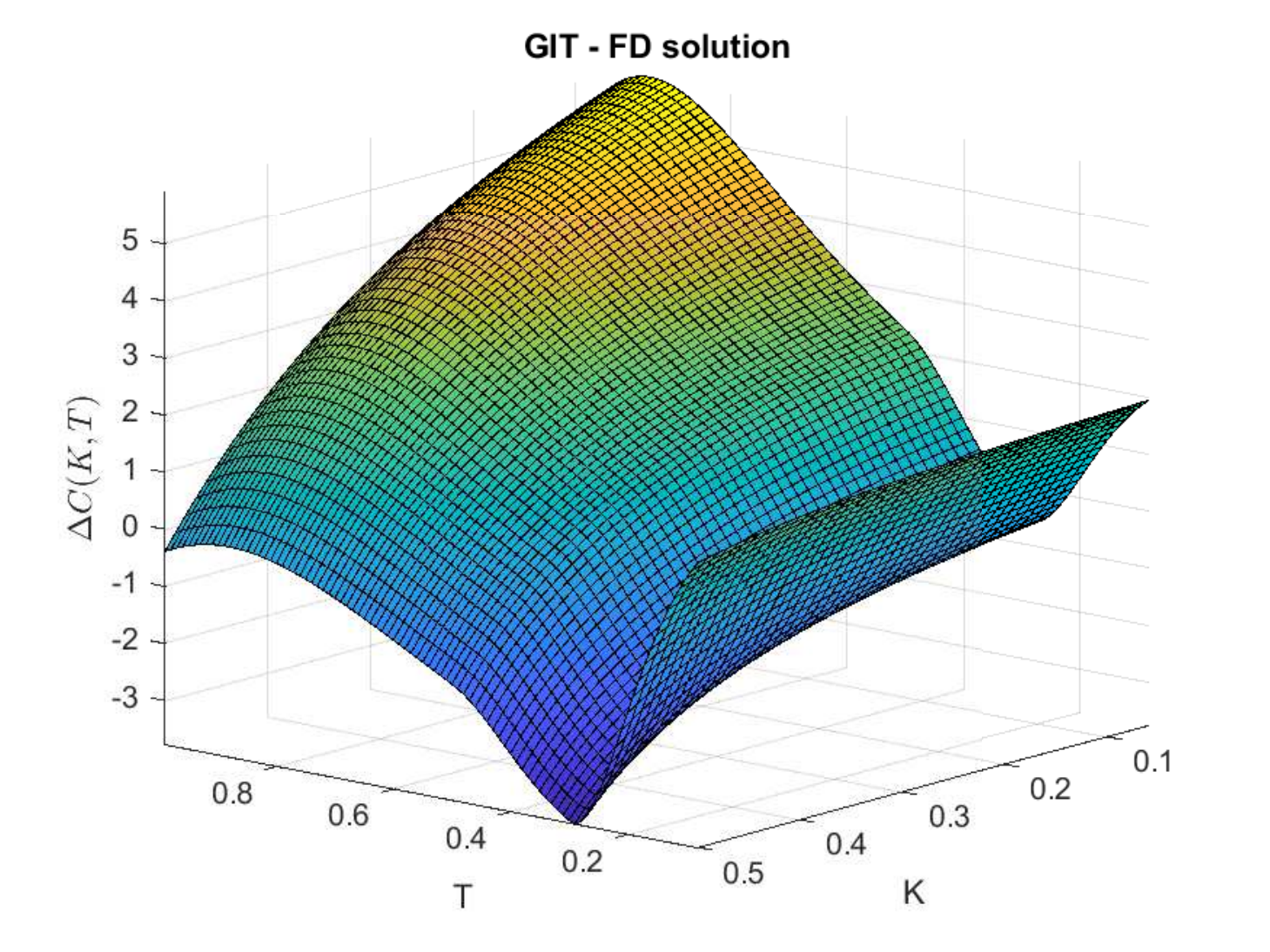}
\caption{The relative difference in \% of the Down-and-Out barrier Call option prices computed by using  the GIT method and the FD solver.}
\label{errorGIT}
\end{figure}

Similar results obtained by using the GIT method are presented in Fig.~\ref{errorGIT}. Computations are done in the same way as for the HP method. The Volterra equation in \eqref{Fin} is solved by discretizing the kernel of the first integral using the trapezoidal rule. The second integral can be computed analytically in terms of Erf functions, similar to how this is done in \eqref{secI}. Actually, first one can compute the second integral in \eqref{u_explicit}, then take the derivative of the result on $x$, and the let $x = y(\tau)$. Overall, we obtain a system of linear equations with respect to the discrete values of $\Upsilon(\tau)$ which matrix is lower triangular with ones on the main diagonal. Therefore, solving this system is trivial. Once the vector $\Upsilon(\tau)$ is found, it could be substituted into \eqref{u_explicit}. Then the final solution is obtained by computing the first integral since the second integral is known analytically.

As mentioned in \citep{CarrItkin2020}, the matrix of the linear system doesn't depend on $K$, just the RHS of the equation. Therefore, the function $\Upsilon(\tau)$ can be found simultaneously for all strikes by using vector elimination (this is equivalent to the solution of the system of linear equations with multiple RHS parts, while the matrix of the system is still lower triangular).

It turns out that the performance of both the GIT and HP methods is same.  However, the GIT method produces more accurate results at high strikes and maturities (i.e. where the option price is relatively small) in contrast to the HP method which is more accurate at short maturities and low strikes. This behavior was explained in detail in \citep{CarrItkinMuravey2020} for the CIR and CEV models. It can be verified by looking at the exponents in Eq. (94) of \citep{CarrItkinMuravey2020} which are proportional to the time $\tau$. Contrary, when the price is higher (short maturities, low strikes) the GIT method is slightly less accurate than the HP method, as in Eq. (45) of \citep{CarrItkinMuravey2020} the exponent is inversely proportional to $\tau$.

However, here the situation is a bit different. Indeed, the final PDE in \citep{CarrItkinMuravey2020} was reduced to the Bessel PDE, why here we work with the heat equation. Therefore, the exponents in both the HP and GIT integrals are inversely proportional to $\tau$.  However, the GIT integrals contain a difference of two exponents which becomes small at large $\tau$, in contrast to the HP exponent which tends to 1. Therefore, the convergence properties of two methods are different at large $\tau$.

This situation is well known for the heat equation with constant coefficients. There exist two representation of the solution: one - obtained by using the method of images, and the other one - by the Fourier series. Despite both solutions are equal as the infinite series, their convergence properties are different. Thus, both GIT and HP methods are complementary.

Also, as observed in our experiments, at high maturities and strikes the first integral in \eqref{u_explicit} is small as compared with the second one. Then the approximate solution is given by the second integral which can be computed analytically. So in this case the whole approximation becomes pure analytical.

\section{Discussion}

In Sections~\ref{GITmethod},\ref{secSol}  we constructed semi-closed form solutions for the prices of Down-and-Out barrier Call option $C_{dao}$ where the underlying is a Zero-Coupon Bond, and the interest rate dynamics follows the Hull-White model. Obviously, using the parity for barrier options, the price of the Down-and-In barrier Call option $C_{dai}$ can be found as $C_{dai} = C_{van} - C_{dao}$, where $C_{van}$ is the price of the European vanilla Call option in the Hull-White model. Since this model allows closed-form solutions for European options on Zero-coupon bonds, \citep{andersen2010interest}, our solution also provides a closed form solution for $C_{dai}$. For the Up-and-Out barrier Call option $C_{uao}$ a simple change of variables $x \to -x$ reduces the pricing problem to that one considered in this paper. Accordingly, the price of the Up-and-In barrier Call option can be found by using the barrier parity. The Puts are priced in the same way.

 From the computational point of view the proposed solution is very efficient as this is shown in Section~\ref{numRes}. Using theoretical analysis justified by a test example we conclude that our method is, at least, of the same complexity, or even faster than the forward FD method. On the other hand, our approach provides high accuracy in computing the options prices, as this is regulated by quadrature rule used to discretize the kernel. Therefore,  the accuracy of the method in $x$ space can be easily increased by using high order quadratures.  For instance, using the Simpson instead of the trapezoid rule doesn't affect the complexity of our method but increase the accuracy, while increasing the accuracy for the FD method is not easy (i.e., it significantly increases the complexity of the method, e.g., see \citep{ItkinBook}).

Also, as mentioned in \citep{CarrItkinMuravey2020}, another advantage of the approach advocated in this paper is computation of option Greeks. Since the option prices in both the HP and GIT methods are represented in closed form via integrals, the explicit dependence of prices on the model parameters is available and transparent. Therefore, explicit representations of the option Greeks can be obtained by a simple differentiation under the integrals. This means that the values of Greeks can be calculated simultaneously with the prices almost with no increase in time. This is because differentiation under the integrals slightly changes the integrands, and these changes could be represented as changes in weights of the quadrature scheme used to numerically compute the integrals. Since the major computational time must be spent for computation of densities which contain special functions, they can be saved during the calculation of the prices, and then reused for computation of Greeks.

\section*{Acknowledgments}

We are grateful to Peter Carr and Alex Lipton for some helpful comments. Dmitry Muravey acknowledges support by the Russian Science Foundation under the Grant number 20-68-47030.

%%%%%%%%%%%%%%%%%%%%%%%%%%%%%%%%%%%%%%%%%%%%%%%%%%%%%%%%%%%%
\vspace{0.4in}
%\printbibliography[title={References}]
%\bibliographystyle{plainnat}
%\bibliography{tdOU}

\end{document}